\pdfoutput=1

\RequirePackage{fix-cm}

\documentclass[twocolumn]{svjour3}

\smartqed

\usepackage{graphicx,amssymb,amsmath}
\usepackage{epsf,psfig}
\usepackage{txfonts}
\usepackage[numbers]{natbib}
\usepackage[unicode]{hyperref}

\journalname{Nonlinear Dynamics}

\begin{document}

\title{Fugitive stars in active galaxies}

\author{Euaggelos E. Zotos}

\institute{Department of Physics, School of Science, \\
Aristotle University of Thessaloniki, \\
GR-541 24, Thessaloniki, Greece \\
Corresponding author's email: {evzotos@physics.auth.gr}
}

\date{Received: 31 July 2015 / Accepted: 23 September 2015 / Published online: 1 October 2015}

\titlerunning{Fugitive stars in active galaxies}

\authorrunning{Euaggelos E. Zotos}

\maketitle

\begin{abstract}

We investigate in detail the escape dynamics in an analytical gravitational model which describes the motion of stars in a quasar galaxy with a disk and a massive nucleus. We conduct a thorough numerical analysis distinguishing between regular and chaotic orbits as well as between trapped and escaping orbits, considering only unbounded motion for several energy levels. In order to distinguish safely and with certainty between ordered and chaotic motion we apply the Smaller ALingment Index (SALI) method. It is of particular interest to locate the escape basins through the openings around the collinear Lagrangian points $L_1$ and $L_2$ and relate them with the corresponding spatial distribution of the escape times of the orbits. Our exploration takes place both in the configuration $(x,y)$ and in the phase $(x,\dot{x})$ space in order to elucidate the escape process as well as the overall orbital properties of the galactic system. Our numerical analysis reveals the strong dependence of the properties of the considered escape basins with the total orbital energy, with a remarkable presence of fractal basin boundaries along all the escape regimes. We hope our outcomes to be useful for a further understanding of the escape mechanism in active galaxy models.


\end{abstract}

\section{Introduction}
\label{intro}

Over the years many studies have been devoted on the issue of escaping particles from dynamical systems. Especially the issue of escapes in Hamiltonian systems is directly related to the problem of chaotic scattering which has been an active field of research over the last decades and it still remains open (e.g., [\citealp{BGOB88}, \citealp{JS88}, \citealp{CK92}, \citealp{BST98}, \citealp{ML02}, \citealp{SASL06}, \citealp{SSL07}, \citealp{SS08}, \citealp{SHSL09}, \citealp{SS10}]). The problem of escape is a classical problem in simple Hamiltonian nonlinear systems (e.g., [\citealp{AVS01}, \citealp{AS03}, \citealp{AVS09}, \citealp{BBS08}, \citealp{BSBS12}, \citealp{Z14a}]) as well as in dynamical astronomy (e.g, [\citealp{HB83}, \citealp{BTS96}, \citealp{BST98}, \citealp{dML00}, \citealp{Z12}]). Escaping orbits in the classical Restricted Three-Body Problem (RTBP) is another typical example (e.g., [\citealp{N04}, \citealp{N05}, \citealp{dAT14}, \citealp{Z15b}]).

Nevertheless, the issue of escaping orbits in Hamiltonian systems is by far less explored than the closely related problem of chaotic scattering. In this situation, a test particle coming from infinity approaches and then scatters off a complex potential. This phenomenon is well investigated as well interpreted from the viewpoint of chaos theory (e.g., [\citealp{BGOB88}, \citealp{BGO90}, \citealp{BOG89}, \citealp{J87}, \citealp{JLS99}, \citealp{JMS95}, \citealp{JP89}, \citealp{JR90}, \citealp{JS87}, \citealp{JT91}, \citealp{LMG00}, \citealp{LGB93}, \citealp{LFO91}, \citealp{LJ99}]).

In open Hamiltonian systems an issue of paramount importance is the determination of the basins of escape, similar to basins of attraction in dissipative systems or even the Newton-Raphson fractal structures. An escape basin is defined as a local set of initial conditions of orbits for which the test particles escape through a certain exit in the Zero Velocity Curve (ZVC) for energies above the escape value. Basins of escape have been studied in many earlier papers (e.g., [\citealp{BGOB88}, \citealp{C02}, \citealp{KY91}, \citealp{PCOG96}]). The reader can find more details regarding basins of escape in [\citealp{C02}], while the review [\citealp{Z14b}] provides information about the escape properties of orbits in a multi-channel dynamical system composed of a two-dimensional perturbed harmonic oscillator. The boundaries of an escape basin may be fractal (e.g., [\citealp{AVS09}, \citealp{BGOB88}]) or even respect the more restrictive Wada property (e.g., [\citealp{AVS01}]), in the case where three or more escape channels coexist in the ZVC.

Escaping and trapped motion of stars in stellar systems is an another issue of great importance. In a previous article [\citealp{Z12}], we explored the nature of orbits of stars in a galactic-type potential, which can be considered to describe local motion in the meridional $(R,z)$ plane near the central parts of an axially symmetric galaxy. It was observed, that apart from the trapped orbits there are two types of escaping orbits, those which escape fast and those which need to spend vast time intervals inside the ZVC before they find the exit and eventually escape. The escape dynamics and the dissolution process of a star cluster embedded in the tidal field of a parent galaxy was investigated in [\citealp{EJSP08}]. Conducting a thorough scanning of the available phase space the authors managed to obtain the basins of escape and the respective escape rates of the orbits, revealing that the higher escape times correspond to initial conditions of orbits near the fractal basin boundaries. The investigation was expanded into three dimensions in [\citealp{Z15a}] where we revealed the escape mechanism of three-dimensional orbits in a tidally limited star cluster. Furthermore, [\citealp{EP14}] explored the escape dynamics in the close vicinity of and within the critical area in a two-dimensional barred galaxy potential, identifying the escape basins both in the phase and the configuration space. The numerical approach of the above-mentioned papers served as the basis of this work.

In an earlier paper [\citealp{PC05}] (hereafter Paper I) the authors introduced a composite dynamical model for the description of the properties of an active galaxy hosting a quasar in its core. In the present paper we shall adopt this analytical gravitational model in order to investigate the escape dynamics of stars. The structure of the paper is as follows: In Section \ref{galmod} we present in detail the properties of our active galaxy model. All the computational methods we used in order to determine the nature of the orbits are described in Section \ref{cometh}. In the following Section, we conduct a thorough and systematic numerical investigation revealing the overall orbital structure (bounded regions and basins of escape) of the active galaxy and showing how it is affected by the total orbital energy. Our paper ends with Section \ref{disc}, where the discussion and the conclusions of this research are given.

\section{Presentation of the galaxy model}
\label{galmod}

The aim of this research is to explore the properties of motion in a quasar model with a disk and a central, massive and spherically symmetric nucleus. For modelling this particular stellar system we use the galaxy model proposed in Paper I.

For the description of the disk we use the following mass-model potential
\begin{equation}
\Phi_{\rm d}(x,y) = \frac{- \ G \ M_{\rm d}}{\sqrt{x^2 + b^2 y^2 + c_{\rm d}^2}},
\label{Vd}
\end{equation}
where $G$ is the gravitational constant, $M_{\rm d}$ is the mass of the nucleus, $b$ is a parameter leading to deviation from axial and spherical symmetry, while $c_{\rm d}$ is the scale length of the disk which acts also as a softening parameter. The generalized anharmonic Plummer potential (\ref{Vd}) has been successfully applied several times in the past in order to realistically model non axially symmetric galaxies (see e.g., [\citealp{CP07}, \citealp{MAB95}, \citealp{PC06}]).

In order to model the central massive nucleus we apply the spherically symmetric Plummer potential [\citealp{P11}]
\begin{equation}
\Phi_{\rm n}(x,y) = \frac{- \ G \ M_{\rm n}}{\sqrt{x^2 + y^2 + c_{\rm n}^2}},
\label{Vn}
\end{equation}
where $M_{\rm n}$ is the mass of the disk, while $c_{\rm n}$ is the scale length of the nucleus.

We consider the case where the central, dense and compact active nucleus of the quasar galaxy rotates clockwise at a constant angular velocity $\Omega > 0$. Therefore, the effective potential is given by
\begin{equation}
\Phi_{\rm eff}(x,y) = \Phi_{\rm n}(x,y) + \Phi_{\rm d}(x,y) - \frac{\Omega^2}{2} \left(x^2 + y^2 \right),
\label{Veff}
\end{equation}
which is in fact a cut through the equatorial plane of the three-dimensional (3D) potential.

In our study, we use a system of galactic units, where the unit of length is 1 kpc, the unit of mass is $2.325 \times 10^7 {\rm M}_\odot$ and the unit of time is $0.9778 \times 10^8$ yr (approximately 100 Myr). The velocity unit is 10 km/s, while $G$ is equal to unity $(G = 1)$. The energy unit (per unit mass) is 100 km$^2$s$^{-2}$. In these units, the values of the involved parameters are: $M_{\rm d} = 5000$ (corresponding to 1.1625 $\times 10^{11}$ ${\rm M}_\odot$), $b^2 = 2.5$, $c_{\rm d} = 1.2$, $M_{\rm n} = 400$ (corresponding to 9.3 $\times 10^{9}$ ${\rm M}_\odot$), $c_{\rm n} = 0.25$, $\Omega = 1.25$ and they remain constant throughout the numerical analysis.

\begin{figure}[!tH]
\includegraphics[width=\hsize]{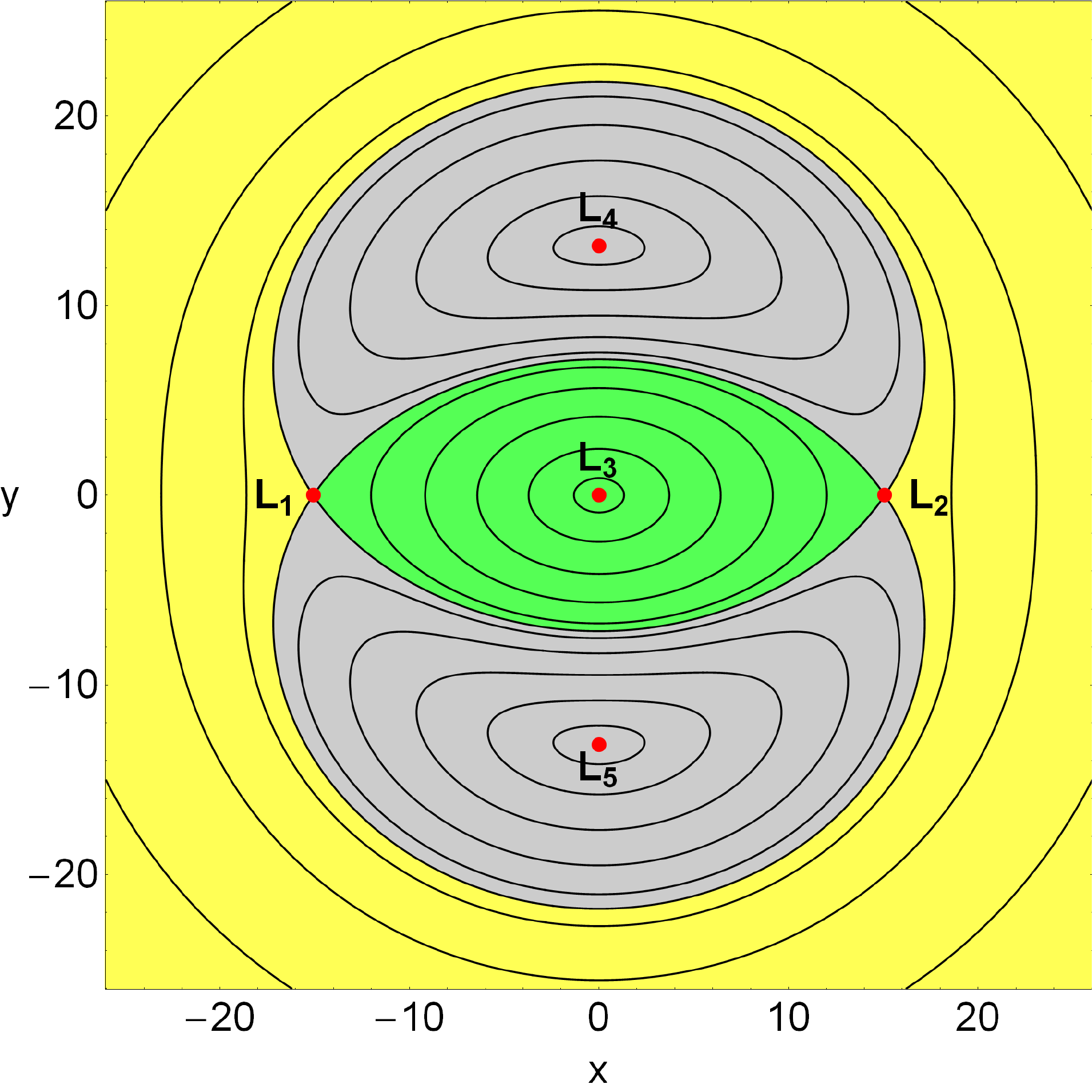}
\caption{The isolines contours of the constant effective potential and the position of the five Lagrangian points. The escaping orbits leak out through exit channels 1 and 2 of the ZVC passing either through $L_1$ or $L_2$, respectively. The interior region is indicated in green, the exterior region is shown in amber color, while the forbidden regions of motion are marked with grey.}
\label{conts}
\end{figure}

This quasar galaxy model has five equilibria called Lagrangian points [\citealp{S67}] at which
\begin{equation}
\frac{\partial \Phi_{\rm eff}}{\partial x} = \frac{\partial \Phi_{\rm eff}}{\partial y} = 0.
\label{lps}
\end{equation}
The isoline contours of constant effective potential as well as the position of the five Lagrangian points $L_i, \ i = {1,5}$ are shown in Fig. \ref{conts}. Three of them, $L_1$, $L_2$, and $L_3$, known as the collinear points, are located on the $x$-axis. The central stationary point $L_3$ at $(x,y) = (0,0)$ is a local minimum of $\Phi_{\rm eff}$. At the other four Lagrangian points it is possible for the test particle to move in a circular orbit, while appearing to be stationary in the rotating frame. For this circular orbit, the centrifugal and the gravitational forces precisely balance. The stationary points $L_1$ and $L_2$ at $(x,y) = (\pm r_L,0) = (\pm 15.07476382858763, 0)$ are saddle points, where $r_L$ is called Lagrangian radius. Let $L_1$ located at $x = -r_L$, while $L_2$ be at $x = +r_L$. The points $L_4$ and $L_5$ on the other hand, which are located at $(x,y) = (0, \pm 13.14171573337381)$ are local maxima of the effective potential, enclosed by the banana-shaped isolines. The annulus bounded by the circles through $L_1$, $L_2$ and $L_4$, $L_5$ is known as the ``region of coroation" (see also [\citealp{BT08}]).

The equations of motion in the rotating frame are described in vectorial form by
\begin{equation}
\ddot{{\vec{r}}} = - {\vec{\nabla}} \Phi_{\rm eff} - 2\left({\vec{\Omega}} \times \dot{{\vec{r}}} \right),
\label{eqmot0}
\end{equation}
where $\vec{r} = (x,y,z)$ is the position vector, $\vec{\Omega} = (0,0,\Omega)$ is the constant rotation velocity vector around the vertical $z$-axis, while the term $- 2\left(\vec{\Omega} \times \dot{\vec{r}} \right)$ represents the Coriolis force. Decomposing Eq. (\ref{eqmot0}) into rectangular Cartesian coordinates $(x,y)$, we obtain
\begin{eqnarray}
\ddot{x} = - \frac{\partial \Phi_{\rm eff}}{\partial x} + 2\Omega\dot{y}, \nonumber \\
\ddot{y} = - \frac{\partial \Phi_{\rm eff}}{\partial y} - 2\Omega\dot{x},
\label{eqmot}
\end{eqnarray}
where the dot indicates derivative with respect to the time.

In the same vein, the equations describing the evolution of a deviation vector ${\vec{w}} = (\delta x, \delta y, \delta \dot{x}, \delta \dot{y})$ which joins the corresponding phase space points of two initially nearby orbits, needed for the calculation of standard chaos indicators (the SALI in our case) are given by the following variational equations
\begin{eqnarray}
\dot{(\delta x)} &=& \delta \dot{x}, \nonumber \\
\dot{(\delta y)} &=& \delta \dot{y}, \nonumber \\
(\dot{\delta \dot{x}}) &=&
- \frac{\partial^2 \Phi_{\rm eff}}{\partial x^2} \ \delta x
- \frac{\partial^2 \Phi_{\rm eff}}{\partial x \partial y} \ \delta y + 2\Omega \ \delta \dot{y},
\nonumber \\
(\dot{\delta \dot{y}}) &=&
- \frac{\partial^2 \Phi_{\rm eff}}{\partial y \partial x} \ \delta x
- \frac{\partial^2 \Phi_{\rm eff}}{\partial y^2} \ \delta y - 2\Omega  \ \delta \dot{x}.
\label{vareq}
\end{eqnarray}

Consequently, the corresponding Hamiltonian (known also as the Jacobian) to the total potential given in Eq. (\ref{Veff}) reads
\begin{equation}
H_{\rm J}(x,y,\dot{x},\dot{y}) = \frac{1}{2} \left(\dot{x}^2 + \dot{y}^2 \right) + \Phi_{\rm eff}(x,y) = E,
\label{ham}
\end{equation}
where $\dot{x}$ and $\dot{y}$ are velocities, while $E$ is the numerical value of the Jacobian, which is conserved since it is an isolating integral of motion. Thus, an orbit with a given value for it's energy integral is restricted in its motion to regions in which $E \leq \Phi_{\rm eff}$, while all other regions are forbidden to the star. The numerical value of the total potential at the two Lagrangian points $L_1$ and $L_2$, that is $\Phi_{\rm eff}(r_L,0)$ and $\Phi_{\rm eff}(-r_L,0)$, respectively yields to a critical Jacobi constant which is equal to $C_L = -534.7029065086583$. This value can be used to define a dimensionless energy parameter as
\begin{equation}
\widehat{C} = \frac{C_L - E}{C_L},
\label{chat}
\end{equation}
where $E$ is some other value of the Jacobian. The dimensionless energy parameter $\widehat{C}$ makes the reference to energy levels more convenient. For $E = C_L$ the ZVC encloses the critical area\footnote{In three dimensions the last closed ZVC of the total potential passing through the Lagrangian points $L_1$ and $L_2$ encloses a critical volume.}, while for a Jacobian value $E > C_L$, or in other words when $\widehat{C} > 0$, the ZVC is open and consequently stars can escape from the system. In Fig. \ref{conts} we observe the two openings (exit channels) near $L_1$ and $L_2$ through which the stars can leak out. In fact, we may say that these two exits act as hoses connecting the interior region (green color) of the dynamical system where $-r_L < x < r_L$ with the ``outside world" of the exterior region (amber color). Exit channel 1 at the negative $x$-direction corresponds to escape through Lagrangian point $L_1$, while exit channel 2 at the positive $x$-direction indicates escape through $L_2$. The forbidden regions of motion within the banana-shaped isolines around $L_4$ and $L_5$ points are shown in Fig. \ref{conts} with gray. It is evident, that for $E < C_L$ the two forbidden regions merge together thus escape is impossible since the interior region cannot communicate with the exterior area.

\section{Computational methods}
\label{cometh}

For investigating the escape dynamics of stars in the quasar galaxy model, we need to define samples of initial conditions of orbits whose properties (bounded or escaping motion) will be identified. For this purpose we define for each value of the total orbital energy (all tested energy levels are above the escape energy), dense uniform grids of $1024 \times 1024$ initial conditions regularly distributed in the area allowed by the value of the energy. Following a typical approach, all orbits are launched with initial conditions inside the Lagrangian radius $(x_0^2 + y_0^2 \leq r_L^2)$, or in other words inside the interior green region shown in Fig. \ref{conts} which is the scattering region in our case. Our investigation takes place both in the configuration $(x,y)$ and in the phase $(x,\dot{x})$ space for a better understanding of the escape process. Furthermore, the grids of the initial conditions of the orbits whose properties will be examined are defined as follows: For the configuration $(x,y)$ space we consider orbits with initial conditions $(x_0, y_0)$ with $\dot{x_0} = 0$, while the initial value of $\dot{y_0}$ is always obtained from the energy integral (\ref{ham}) as $\dot{y_0} = \dot{y}(x_0,y_0,\dot{x_0},E) > 0$. Similarly, for the phase $(x,\dot{x})$ space we consider orbits with initial conditions $(x_0, \dot{x_0})$ with $y_0 = 0$, while this time the initial value of $\dot{y_0}$ is obtained from the Jacobi integral (\ref{ham}).

The equations of motion as well as the variational equations for the initial conditions of all orbits were integrated using a double precision Bulirsch-Stoer \verb!FORTRAN 77! algorithm (e.g., [\citealp{PTVF92}]) with a small time step of order of $10^{-2}$, which is sufficient enough for the desired accuracy of our computations. Here we should emphasize, that our previous numerical experience suggests that the Bulirsch-Stoer integrator is both faster and more accurate than a double precision Runge-Kutta-Fehlberg algorithm of order 7 with Cash-Karp coefficients (e.g., [\citealp{DMCG12}]). Throughout all our computations, the Jacobian energy integral (Eq. (\ref{ham})) was conserved better than one part in $10^{-11}$, although for most orbits it was better than one part in $10^{-12}$.

\begin{figure}[!tH]
\includegraphics[width=\hsize]{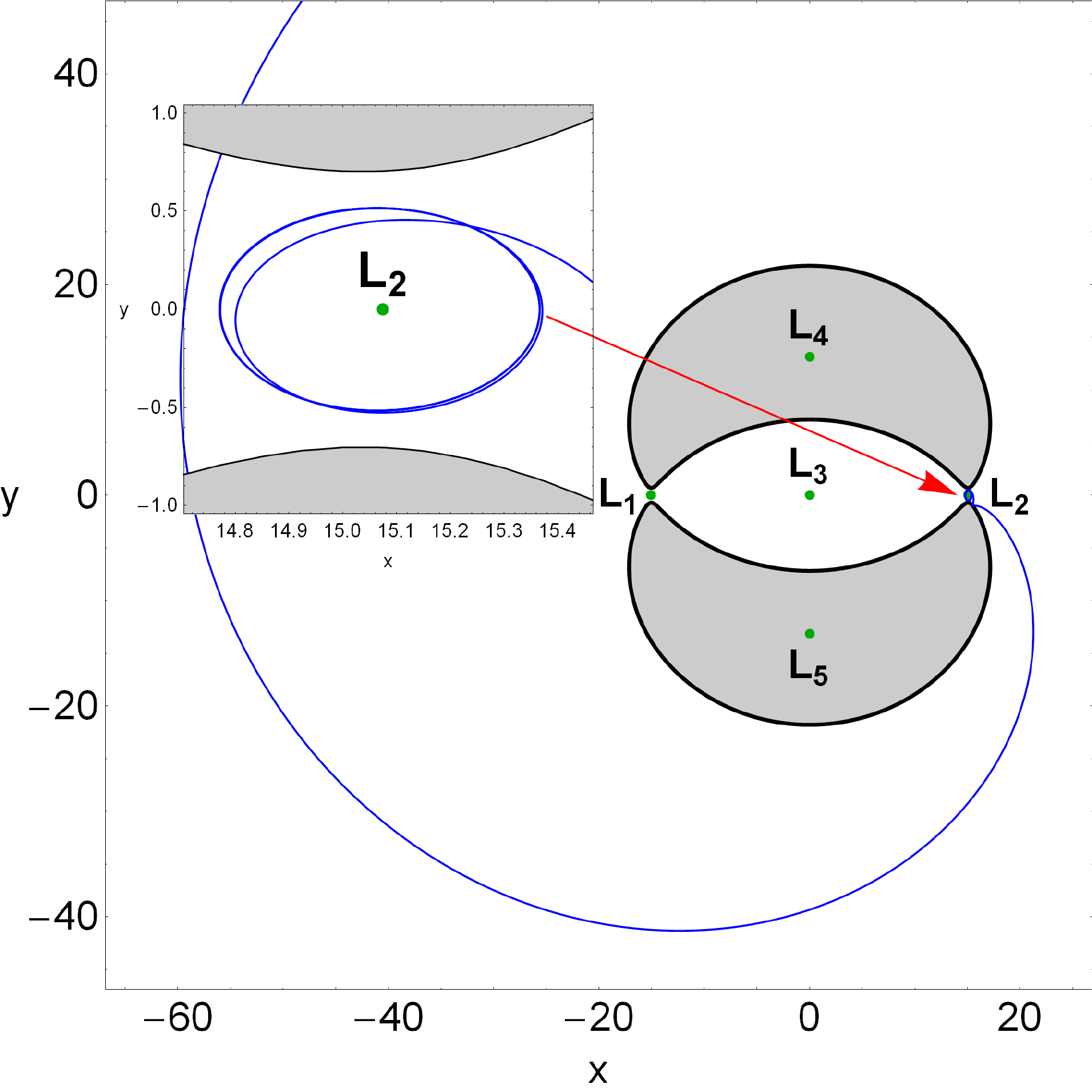}
\caption{The 1:1 highly unstable Lyapunov periodic orbit around $L_2$, when $\widehat{C} = 0.001$. The initial conditions of the orbit are: $x_0 = 14.77108191$, $y_0 = \dot{x_0} = 0$, $\dot{y_0} > 0$. We observe that the orbit performs several loops around the Lagrangian point $L_2$ before it escapes from the system due to its highly instability.}
\label{lyap}
\end{figure}

In dynamical systems with escapes an issue of paramount importance is the determination of the position as well as the time at which an orbit escapes. For all values of energy smaller than the critical value $C_L$ (escape energy), the ZVC is closed and therefore all orbits are bounded. On the other hand, when $E > C_L$ the ZVC is open and extend to infinity. The value of the energy itself however, does not furnish a sufficient condition for escape. An open ZVC consists of two branches forming channels through which an orbit can escape to infinity (see Fig. \ref{conts}). It was proved [\citealp{C79}] that in Hamiltonian systems there is a highly unstable periodic orbit at every opening close to the line of maximum potential which is called a Lyapunov orbit. In fact, there is a family of unstable Lyapunov orbits around each collinear Lagrangian point [\citealp{M58}]. Such an orbit reaches the ZVC\footnote{The boundaries of the accessible regions of motion are the so-called Zero Velocity Curves, since they are the locus in the $(x,y)$ space where kinetic energy vanishes.}, on both sides of the opening and returns along the same path thus, connecting two opposite branches of the ZVC. In Fig. \ref{lyap} we present the shape of the unstable Lyapunov orbit around $L_2$, when $\widehat{C} = 0.001$. It is seen that the Lyapunov orbit eventually escapes from the system after having performed several loops around the Lagrangian point $L_2$. Lyapunov orbits are very important for the escapes from the system, since if an orbit intersects any one of these orbits with velocity pointing outwards moves always outwards and eventually escapes from the system without any further intersections with the surface of section (e.g., [\citealp{C90}]). When $E = C_L$ the Lagrangian points exist precisely but when $E > C_L$ an unstable Lyapunov periodic orbit is located close to each of these two points (e.g., [\citealp{H69}]). These Lagrangian points $L_1$ and $L_2$ are saddle points of the total potential, so when $E > C_L$, a star must pass close enough to one of these points in order to escape. Thus, in our quasar galaxy system the escape criterion is purely geometric. In particular, escapers are defined to be those stars moving in orbits beyond the Lagrangian radius $(r_L)$ thus passing through one of the two Lagrangian points ($L_1$ or $L_2$) and intersecting one of the two unstable Lyapunov orbits with velocity pointing outwards. Here we must emphasize that orbits with initial conditions outside the interior region, or in other words outside $L_1$ or $L_2$, do not necessarily escape immediately from the galaxy. Thus the initial position itself does not furnish a sufficient condition for escape, since the escape criterion is a combination of the coordinates and the velocity of stars.

The configuration and the phase space are divided into the escaping and non-escaping (trapped) space. Usually, the vast majority of the trapped space is occupied by initial conditions of regular orbits forming stability islands where a third integral of motion is present. In many systems however, trapped chaotic orbits have also been observed (see e.g., [\citealp{Z15a}]). Therefore, we decided to distinguish between regular and chaotic trapped motion. Over the years, several chaos indicators have been developed in order to determine the character of orbits. In our case, we chose to use the Smaller ALingment Index (SALI) method. The SALI [\citealp{S01}] has been proved a very fast, reliable and effective tool, which is defined as
\begin{equation}
\rm SALI(t) \equiv min(d_-, d_+),
\label{sali}
\end{equation}
where $d_- \equiv \| {\vec{w_1}}(t) - {\vec{w_2}}(t) \|$ and $d_+ \equiv \| {\vec{w_1}}(t) + {\vec{w_2}}(t) \|$ are the alignments indices, while ${\vec{w_1}}(t)$ and ${\vec{w_2}}(t)$, are two deviation vectors which initially point in two random directions. For distinguishing between ordered and chaotic motion, all we have to do is to compute the SALI along time interval $t_{max}$ of numerical integration. In particular, we track simultaneously the time-evolution of the main orbit itself as well as the two deviation vectors ${\vec{w_1}}(t)$ and ${\vec{w_2}}(t)$ in order to compute the SALI.

The time-evolution of SALI strongly depends on the nature of the computed orbit since when an orbit is regular the SALI exhibits small fluctuations around non zero values, while on the other hand, in the case of chaotic orbits the SALI after a small transient period it tends exponentially to zero approaching the limit of the accuracy of the computer $(10^{-16})$. Therefore, the particular time-evolution of the SALI allows us to distinguish fast and safely between regular and chaotic motion. Nevertheless, we have to define a specific numerical threshold value for determining the transition from order to chaos. After conducting extensive numerical experiments, integrating many sets of orbits, we conclude that a safe threshold value for the SALI is the value $10^{-8}$. Thus, in order to decide whether an orbit is regular or chaotic, one may follow the usual method according to which we check after a certain and predefined time interval of numerical integration, if the value of SALI has become less than the established threshold value. Therefore, if SALI $\leq 10^{-8}$ the orbit is chaotic, while if SALI $ > 10^{-8}$ the orbit is regular thus making the distinction between regular and chaotic motion clear and beyond any doubt. For the computation of SALI we used the \verb!LP-VI! code [\citealp{CMD14}], a fully operational routine which efficiently computes a suite of many chaos indicators for dynamical systems in any number of dimensions.

In our computations, we set $10^4$ time units as a maximum time of numerical integration. The vast majority of orbits (regular and chaotic) however, need considerable less time to find one of the two exits in the ZVC and eventually escape from the system (obviously, the numerical integration is effectively ended when an orbit passes through one of the escape channels and escapes). Nevertheless, we decided to use such a vast integration time just to be sure that all orbits have enough time in order to escape. Remember, that there are the so called ``sticky orbits" which behave as regular ones during long periods of time. Here we should clarify, that orbits which do not escape after a numerical integration of $10^4$ time units are considered as non-escaping or trapped. In fact, orbits with escape periods equal to many Hubble times are completely irrelevant to our investigation since they lack physical meaning.

\section{Numerical results}
\label{numres}

\begin{figure*}[!tH]
\centering
\resizebox{0.8\hsize}{!}{\includegraphics{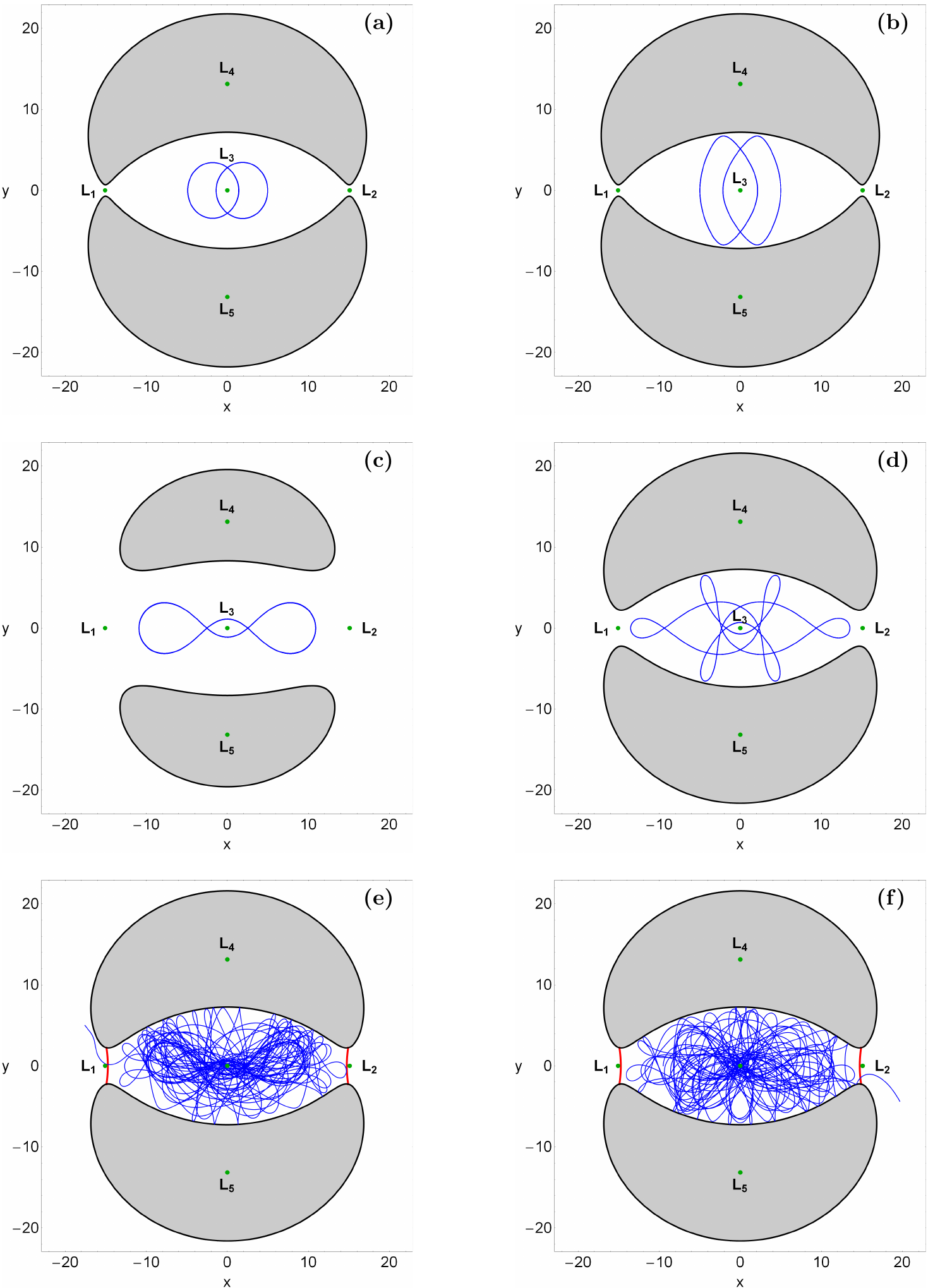}}
\caption{Characteristic examples of the main types of orbits in our quasar galaxy model. (a-upper left): two 1:1 retrograde loop orbits, (b-upper right): two 1:1 prograde loop orbits, (c-middle left): a 1:3 resonant quasi-periodic orbit, (d-middle right): a 3:7 secondary resonant quasi-periodic orbit, (e-lower left): an orbit escaping through $L_1$, (e-lower right): an orbit escaping through $L_2$. The two unstable Lyapunov orbits around the Lagrangian points are visualized in red. More details are given in Table \ref{table1}.}
\label{orbs}
\end{figure*}

The main objective of our exploration is to determine which orbits escape and which remain trapped, distinguishing simultaneously between regular and chaotic trapped motion\footnote{Generally, any dynamical method requires a sufficient time interval of numerical integration in order to distinguish safely between ordered and chaotic motion. Therefore, if the escape rate of an orbit is very low or even worse if the orbit escapes directly from the system then, any chaos indicator (the SALI in our case) will fail to work properly due to insufficient integration time. Hence, it is pointless to speak of regular or chaotic escaping orbits.}. Furthermore, two additional properties of the orbits will be examined: (i) the channels or exits through which the stars escape and (ii) the time-scale of the escapes (we shall also use the terms escape period or escape rates). We will investigate these dynamical quantities for various values of the energy, always within the interval $\widehat{C} \in [0.001,0.1]$.

Our numerical calculations indicate that apart from the escaping orbits there is always a considerable amount of non-escaping orbits. In general terms, the majority of non-escaping regions corresponds to initial conditions of regular orbits, where a third integral of motion is present, restricting their accessible phase space and therefore hinders their escape. However, there are also chaotic orbits which do not escape within the predefined interval of $10^4$ time units and remain trapped for vast periods until they eventually escape to infinity. At this point, it should be emphasized and clarified that these trapped chaotic orbits cannot be considered, by no means, neither as sticky orbits nor as super sticky orbits with sticky periods larger than $10^4$ time units. Sticky orbits are those who behave regularly for long time periods before their true chaotic nature is fully revealed. In our case on the other hand, this type of orbits exhibit chaoticity very quickly as it takes no more than about 100 time units for the SALI to cross the threshold value (SALI $\ll 10^{-8}$), thus identifying beyond any doubt their chaotic character. Therefore, we decided to classify the initial conditions of orbits in both the configuration and phase space into four main categories: (i) orbits that escape through $L_1$, (ii) orbits that escape through $L_2$, (iii) non-escaping regular orbits and (iv) trapped chaotic orbits.

\begin{table}[!ht]
\centering
\setlength{\tabcolsep}{10pt}
\begin{center}
   \centering
   \caption{Type, initial conditions, value of the energy and integration time of the orbits shown in Fig. \ref{orbs}(a-f). For all orbits we have $y_0 = \dot{x_0} = 0$, while the initial value of $\dot{y}$ is obtained from the Jacobi integral (\ref{ham}).}
   \label{table1}
   \begin{tabular}{@{}lccccccr}
      \hline
      Figure & Type & $x_0$ & $\widehat{C}$ & $t_{\rm int}$  \\
      \hline
      \ref{orbs}a & 1:1 loop &  -4.90 & 0.001 & 100 \\
      \ref{orbs}a & 1:1 loop &  -1.40 & 0.001 & 100 \\
      \ref{orbs}b & 1:1 loop &   2.14 & 0.001 & 100 \\
      \ref{orbs}b & 1:1 loop &   4.99 & 0.001 & 100 \\
      \ref{orbs}c & 1:3 box  & -10.90 & 0.100 & 100 \\
      \ref{orbs}d & 3:7 box  & -13.53 & 0.010 & 100 \\
      \ref{orbs}e & chaotic  & -10.00 & 0.010 &  90 \\
      \ref{orbs}f & chaotic  &  -9.50 & 0.010 &  93 \\
      \hline
   \end{tabular}
\end{center}
\end{table}

Additional numerical computations reveal that the non-escaping regular orbits are mainly 1:1 loop orbits for which a third integral of motion applies, while other types of secondary resonant orbits are also present. In Fig. \ref{orbs}(a-b) we present two sets of 1:1 thin loop orbits (retrograde and prograde). Moreover in Fig. \ref{orbs}c we see a typical example of a secondary 1:3 resonant quasi-periodic orbit, while Fig. \ref{orbs}d shows a complicated 3:7 secondary resonant quasi-periodic orbit. The $n:m$ notation we use for the regular orbits is according to [\citealp{CA98}] and [\citealp{ZC13}], where the ratio of those integers corresponds to the ratio of the main frequencies of the orbit, where main frequency is the frequency of greatest amplitude in each coordinate. Main amplitudes, when having a rational ratio, define the resonances of an orbit. Finally in Figs. \ref{orbs}(e-f) we observe two orbits escaping through $L_1$ (exit channel 1) and $L_2$ (exit channel 2), respectively. The regular orbits shown in Figs. \ref{orbs}(a-d) were computed until $t = 100$ time units, while on the other hand, the escaping orbits presented in Figs. \ref{orbs}(e-f) were calculated for 3 time units more than the corresponding escape period in order to visualize better the escape trail. For all orbits we have $y_0 = \dot{x_0} = 0$, while the initial value of $\dot{y}$ is obtained from the Jacobi integral (\ref{ham}). The two unstable Lyapunov orbits around $L_1$ and $L_2$ are shown in Fig. \ref{orbs}(e-f) in red. In Table \ref{table1} we provide the type, the exact initial conditions and the value of the energy for all the depicted orbits.

\subsection{Structure of the configuration $(x,y)$ space}
\label{pp1}

\begin{figure*}[!tH]
\centering
\resizebox{0.8\hsize}{!}{\includegraphics{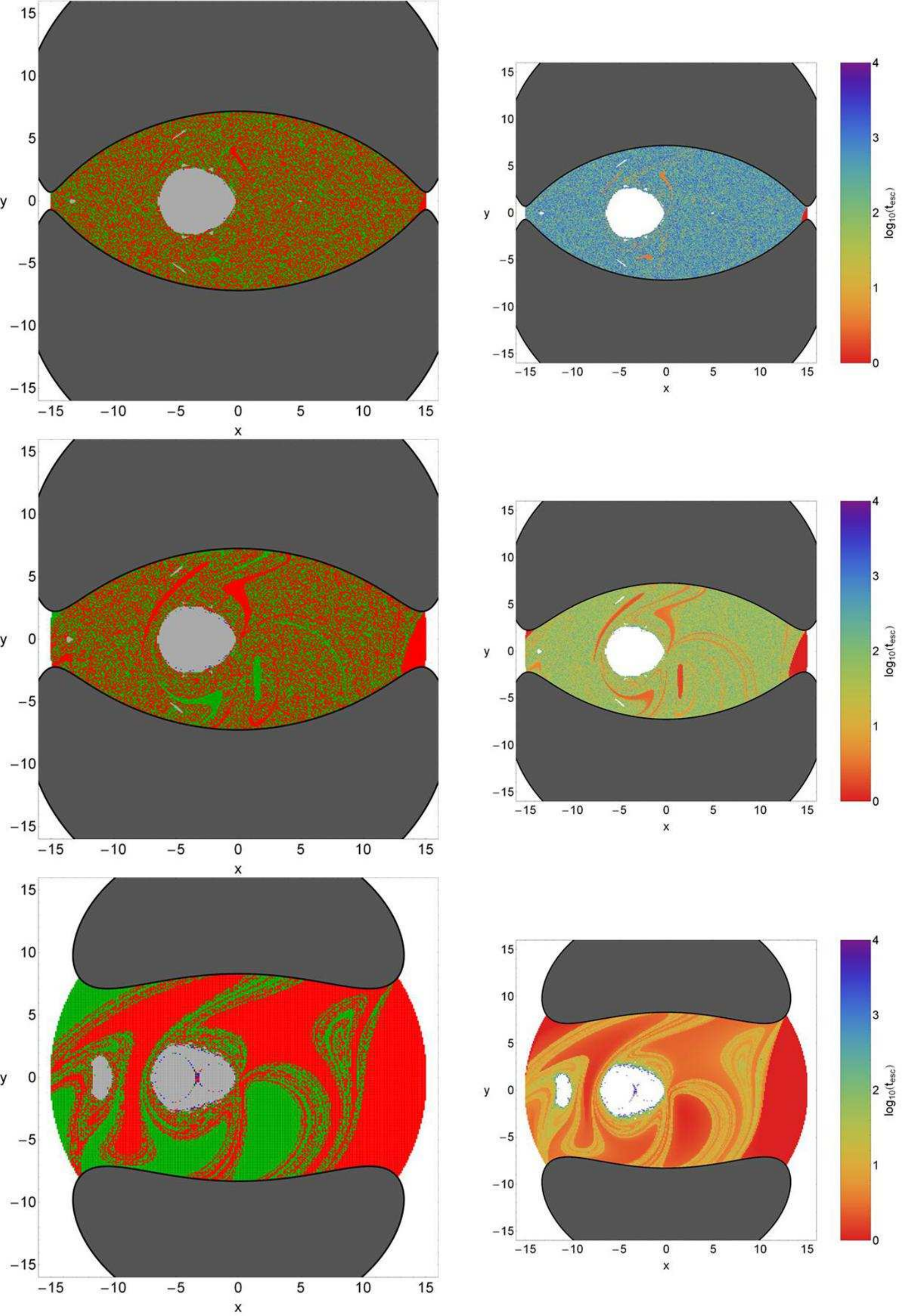}}
\caption{Left column: Orbital structure of the configuration $(x,y)$ space. Top row: $\widehat{C} = 0.001$; Middle row: $\widehat{C} = 0.01$; Bottom row: $\widehat{C} = 0.1$. The green regions correspond to initial conditions of orbits where the stars escape through $L_1$, red regions denote initial conditions where the stars escape through $L_2$, light gray areas represent stability islands of regular non-escaping orbits, initial conditions of trapped chaotic orbits are marked in blue, while the regions of forbidden motion are shown in dark gray. Right column: Distribution of the corresponding escape times $t_{\rm esc}$ of the orbits on the configuration space. The darker the color, the larger the escape time. The scale on the color-bar is logarithmic. Initial conditions of non-escaping regular orbits and trapped chaotic orbits are shown in white.}
\label{grd1}
\end{figure*}

Our exploration begins in the configuration space and in the left column of Fig. \ref{grd1} we present the orbital structure of the $(x,y)$ plane for three values of the energy, where the initial conditions of the orbits are classified into four categories by using different colors. Specifically, light gray color corresponds to regular non-escaping orbits, blue color corresponds to trapped chaotic orbits, green color corresponds to orbits escaping through channel 1, while the initial conditions of orbits escaping through exit channel 2 are marked with red color. The outermost black solid line is the ZVC which is defined as $\Phi_{\rm eff}(x,y) = E$. For $\widehat{C} = 0.001$, that is an energy level just above the critical escape energy $C_L$, we see that the vast majority of initial conditions corresponds to escaping orbits however, a main stability island of non-escaping regular orbits is present denoting ordered motion. We also see that the entire interior region is completely fractal which means that there is a highly dependence of the escape mechanism on the particular initial conditions of the orbits. In other words, a minor change in the initial conditions has as a result the star to escape through the opposite exit channel, which is of course, a classical indication of chaotic motion. With a much closer look at the $(x,y)$ plane we can identify some additional tiny stability islands which are deeply buried in the escape domain corresponding to secondary resonant orbits (see Fig. \ref{orbs}d). As we proceed to higher energy levels the fractal area reduces and several basins of escape start to emerge. By the term basin of escape, we refer to a local set of initial conditions that corresponds to a certain escape channel. Indeed for $\widehat{C} = 0.01$ we observe the presence of escape basins which mainly have the shape of thin elongated bands. Moreover, the extent of the stability islands seems to be unaffected by the increase in the orbital energy. With increasing energy the interior region in the configuration space becomes less and less fractal and broad, well-defined basins of escape dominate when $\widehat{C} = 0.1$. The fractal regions on the other hand, are confined mainly near the boundaries between the escape basins or in the vicinity of the stability islands. Furthermore, it is seen that for $\widehat{C} = 0.1$ two stability islands are present at the left part of the center of the galaxy, while all smaller stability regions have disappeared. We should mention the existence of a figure-eight separatrix inside the main stability island which is mainly composed of initial conditions of trapped chaotic orbits. Our classification indicates that in all three energy levels trapped chaotic motion is almost negligible as the initial conditions of such orbits mainly appear only as lonely points around the boundaries of the stability islands. Here we must point out two interesting phenomena that take place with increasing energy: (i) the regions of forbidden motion are significantly confined, (ii) the two escape channels become more and more wide.

\begin{figure}[!tH]
\includegraphics[width=\hsize]{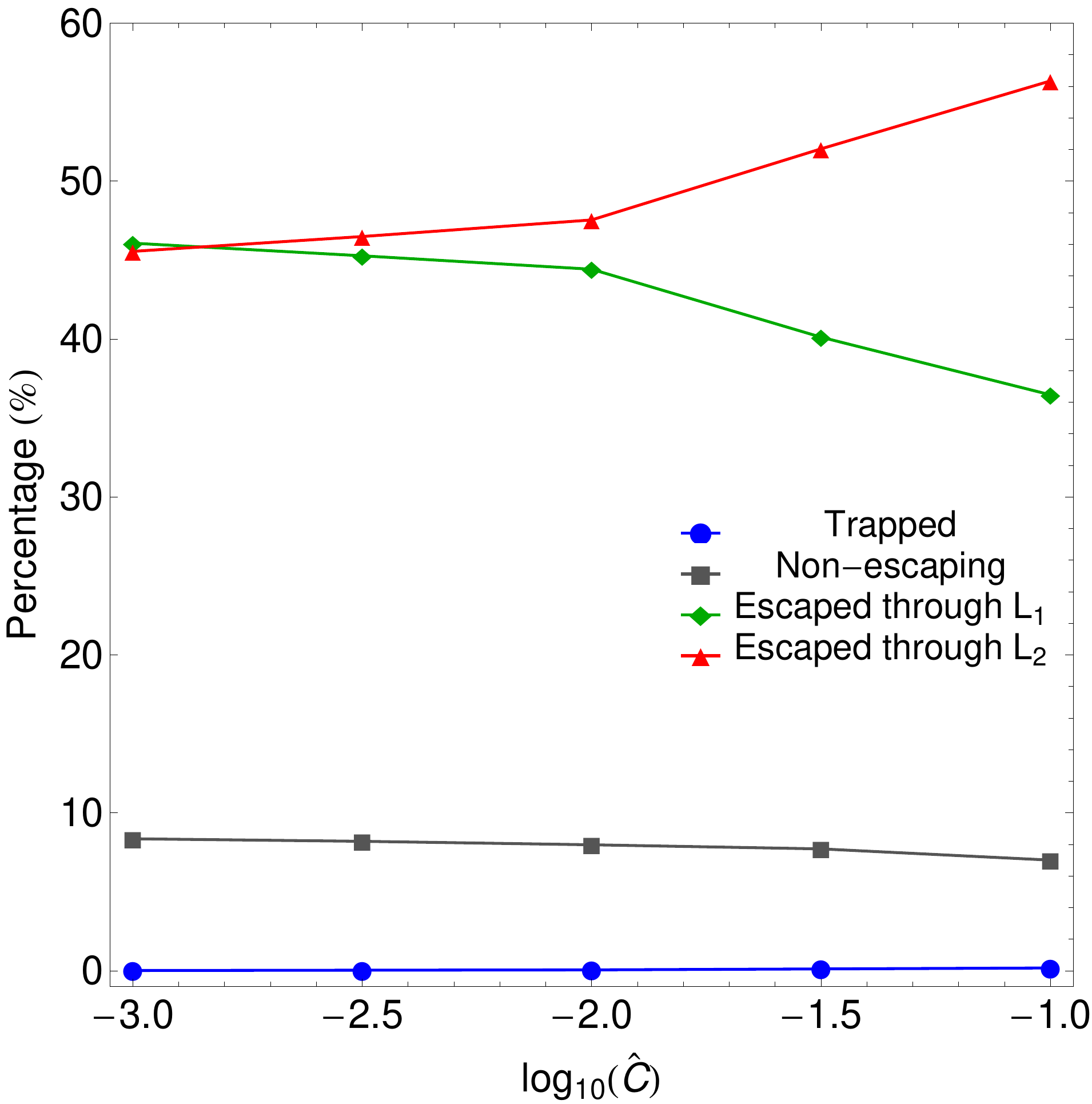}
\caption{Evolution of the percentages of trapped, escaping and non-escaping orbits on the configuration $(x,y)$ space when varying the dimensionless energy parameter $\widehat{C}$.}
\label{percs1}
\end{figure}

In the right column of Fig. \ref{grd1} we show how the escape times $t_{\rm esc}$ of orbits are distributed on the configuration $(x,y)$ space. The escape time $t_{\rm esc}$ is defined as the time when a star crosses one of the Lyapunov orbits with velocity pointing outwards. Light reddish colors correspond to fast escaping orbits with short escape periods, dark blue/purple colors indicate large escape rates, while white color denote both non-escaping regular and trapped chaotic orbits. The scale on the color bar is logarithmic. It is evident, that orbits with initial conditions close to the boundaries of the stability islands need significant amount of time in order to escape from the galaxy, while on the other hand, inside the basins of escape where there is no dependence on the initial conditions whatsoever, we measured the shortest escape rates of the orbits. We observe that for $\widehat{C} = 0.001$ the escape periods of orbits with initial conditions in the fractal region are huge corresponding to tens of thousands of time units. As the value of the total orbital energy increases however, the escape times of orbits reduce significantly. In fact for $\widehat{C} = 0.01$ and $\widehat{C} = 0.1$ the basins of escape can be distinguished in the grids of Fig. \ref{grd1}, being the regions with reddish colors indicating extremely fast escaping orbits. Our numerical calculations indicate that orbits with initial conditions inside the basins have significantly small escape periods of less than 10 time units.

The following Fig. \ref{percs1} presents the evolution of the percentages of the different types of orbits on the configuration $(x,y)$ space when the dimensionless energy parameter $\widehat{C}$ varies. One may observe, that when $\widehat{C} = 0.001$, that is just above the critical escape energy $C_L$, escaping orbits through $L_1$ and $L_2$ share about 90\% of the available configuration space, so the two exit channels can be considered equiprobable since the configuration space is highly fractal. As the value of the energy increases however, the percentages of escaping orbits through exit channels 1 and 2 start to diverge, especially for $\widehat{C} > 0.01$. In particular, the amount of orbits that escape through $L_1$ exhibits a linear decrease, while on the other hand, the rate of escaping orbits through the opposite exit channel displays a linear increase. At the highest energy level studied $(\widehat{C} = 0.1)$ escaping orbits through $L_2$ is the most populated family occupying about 56\% of the configuration plane, while only about 36\% of the same plane corresponds to initial conditions of orbits that escape through $L_1$. Furthermore, the percentage of non-escaping regular orbits is much less affected by the shifting of the energy displaying only a minor decrease form 8.4\% to about 7\%. Finally, the rate of trapped chaotic orbits is always less than 0.2\%. Thus taking into account all the above-mentioned analysis we may conclude that at low energy levels where the fractility of the configuration space is maximum the stars do not show any particular preference regarding the escape channel, while on the contrary, at high enough energy levels where basins of escape dominate it seems that exit channel 2 is more preferable.

\subsection{Structure of the phase $(x,\dot{x})$ space}
\label{pp2}

\begin{figure*}[!tH]
\centering
\resizebox{0.8\hsize}{!}{\includegraphics{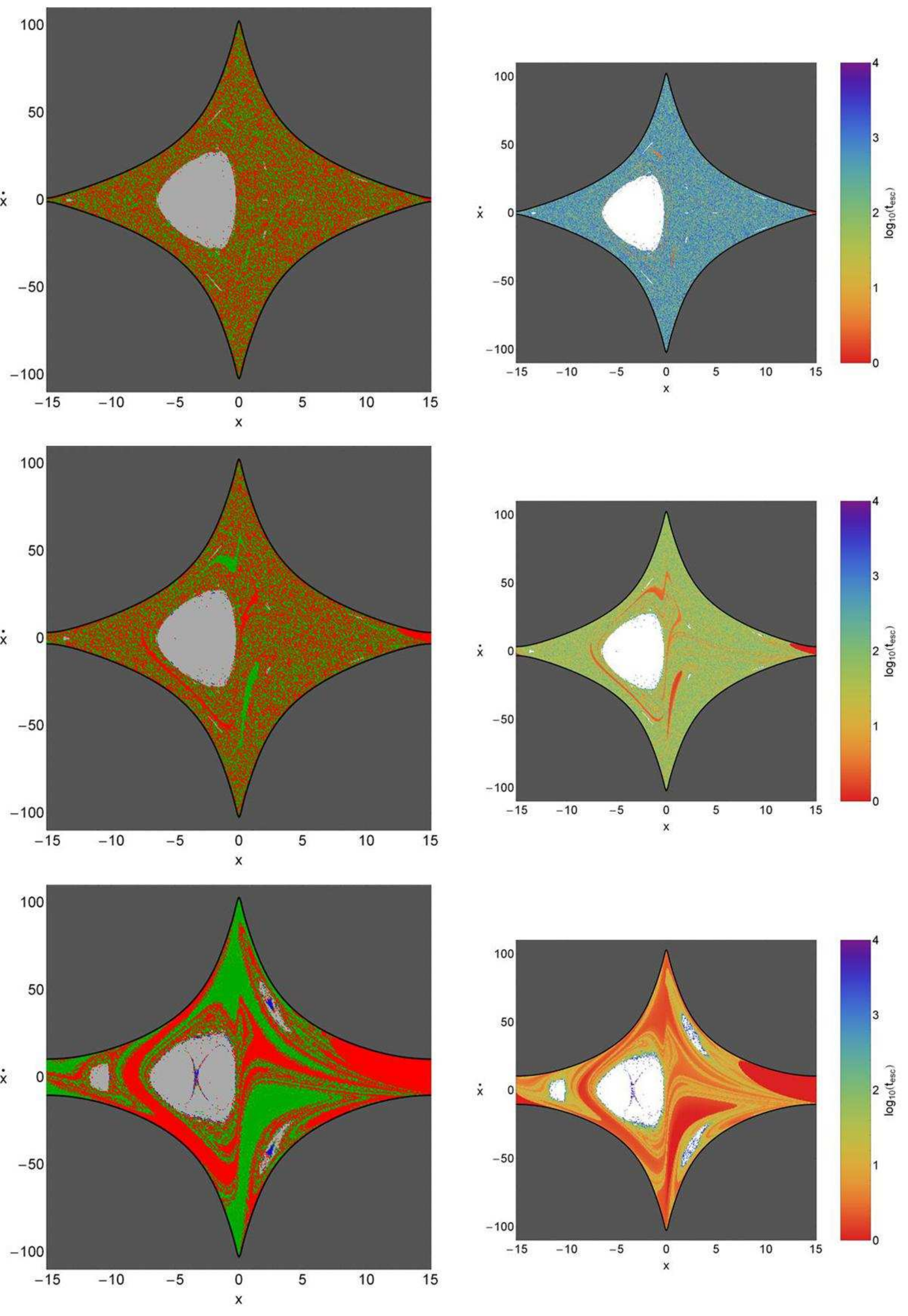}}
\caption{Left column: Orbital structure of the phase $(x,\dot{x})$ space. Top row: $\widehat{C} = 0.001$; Middle row: $\widehat{C} = 0.01$; Bottom row: $\widehat{C} = 0.1$. Right column: Distribution of the corresponding escape times $t_{\rm esc}$ of the orbits on the phase space. The color codes are the same as in Fig. \ref{grd1}.}
\label{grd2}
\end{figure*}

We continue our investigation in the phase $(x,\dot{x})$ space and we follow the same numerical approach as discussed previously. In Fig. \ref{grd2} we depict the orbital structure of the $(x,\dot{x})$ plane for three values of the energy, using different colors in order to distinguish between the four main types of orbits (non-escaping regular; trapped chaotic; escaping through $L_1$ and escaping through $L_2$). The outermost black solid line is the limiting curve which is defined as
\begin{equation}
f(x,\dot{x}) = \frac{1}{2}\dot{x}^2 + \Phi_{\rm eff}(x, y = 0) = E.
\label{zvc}
\end{equation}
Here we must point out, that this $(x,\dot{x})$ phase plane is not a classical Poincar\'{e} Surface of Section (PSS), simply because escaping orbits in general, do not intersect the $y = 0$ axis after a certain time, thus preventing us from defying a recurrent time. A classical Poincar\'{e} surface of section exists only if orbits intersect an axis, like $y = 0$, at least once within a certain time interval. Nevertheless, in the case of escaping orbits we can still define local surfaces of section which help us to understand the orbital behavior of the dynamical system.

We see that just above the escape energy, that is when $\widehat{C} = 0.001$, the structure of the phase plane is very similar to that of the corresponding configuration plane shown in Fig. \ref{grd1}. It is observed that about 85\% of the $(x,\dot{x})$ plane is occupied by initial conditions of escaping orbits. Moreover, the vast escape domain is highly fractal, while basins of escape, if any, are negligible. In all studied cases the areas of regular motion correspond mainly to retrograde orbits (i.e., when a star revolves around one galaxy in the opposite sense with respect to the motion of the galaxy itself), while there are also some smaller stability islands of prograde orbits. The area on the phase plane covered by escape basins grows drastically with increasing energy and at high enough energy levels the fractal regions are confined mainly at the boundaries of the stability islands of non-escaping orbits. In particular, the escape basins first appear at mediocre values of energy $(\widehat{C} = 0.01)$ inside the fractal, extended escape region having the shape of thin elongated bands however, as we proceed to higher energy levels they grow in size dominating the phase space as well-formed broad domains. We observe for $\widehat{C} = 0.1$ the set of the three stability islands corresponding to the 1:3 resonant periodic orbits. The Coriolis forces dictated by the rotation of the central compact nucleus makes the phase planes to be asymmetric with respect to the $\dot{x}$-axis and this phenomenon is usually known as ``Coriolis asymmetry" (e.g., [\citealp{I80}]). The distribution of the corresponding escape times $t_{\rm esc}$ of orbits on the phase space as a function of the energy parameter is shown in the right column Fig. \ref{grd2}. One may observe that the results are very similar to those presented earlier in Fig. \ref{grd1}, where we found that orbits with initial conditions inside the basins of escape have the smallest escape rates, while on the other hand, the longest escape times correspond to orbits with initial conditions either in the fractal regions of the plots, or near the boundaries of the islands of regular motion.

\begin{figure}[!tH]
\includegraphics[width=\hsize]{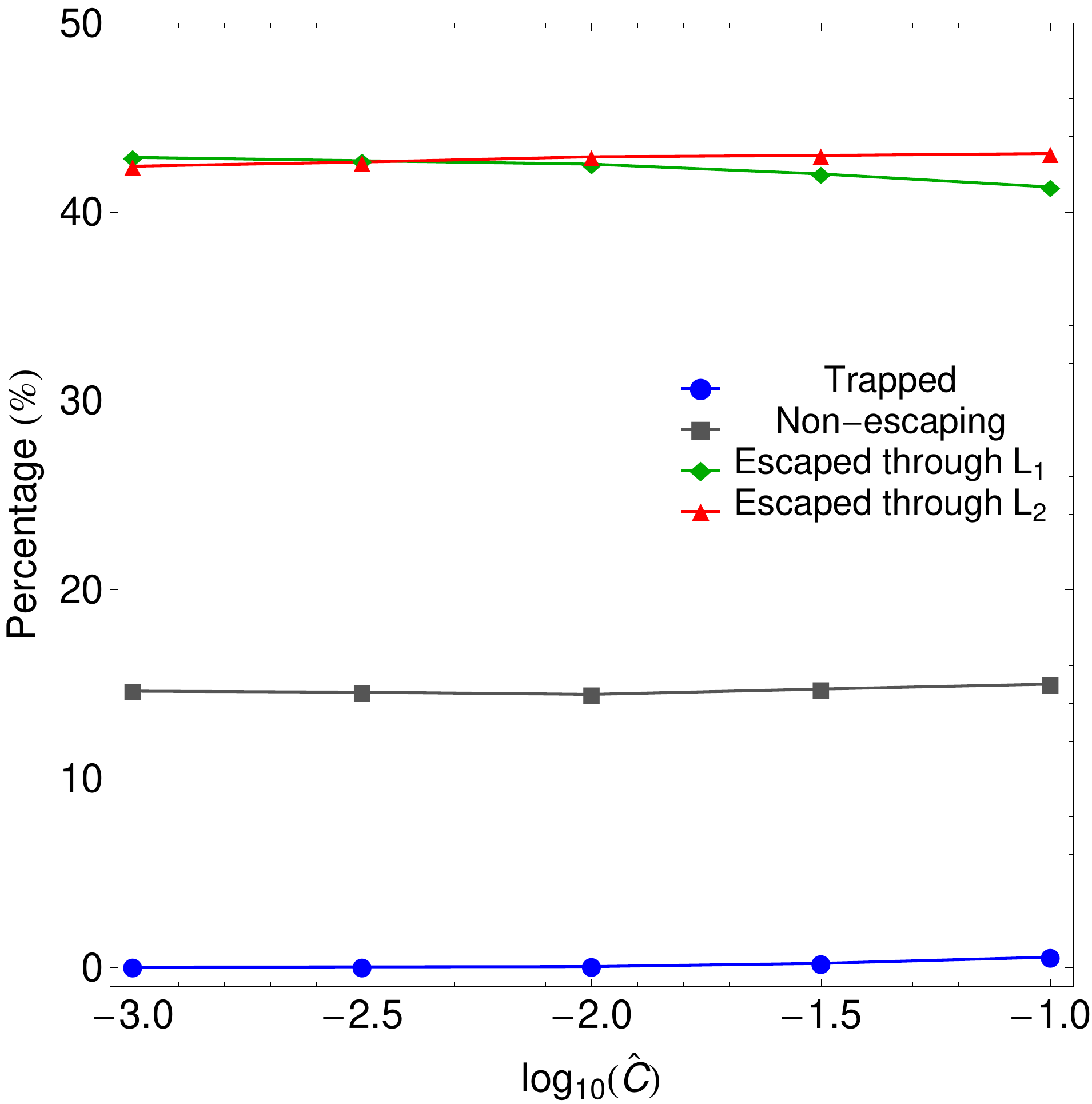}
\caption{Evolution of the percentages of trapped, escaping and non-escaping orbits on the phase $(x,\dot{x})$ space when varying the dimensionless energy parameter $\widehat{C}$.}
\label{percs2}
\end{figure}

The evolution of the percentages of the different types of orbits on the phase $(x,\dot{x})$ space as a function of the dimensionless energy parameter $\widehat{C}$ is presented in Fig. \ref{percs2}. It is seen, that at the lowest energy level studied $(\widehat{C} = 0.001)$ escaping orbits through $L_1$ and $L_2$ share about 85\% of the phase plane, while only 15\% of the same plane corresponds to initial conditions of non-escaping regular orbits. Furthermore, we observe that for $0.001 < \widehat{C} < 0.01$ the rates of all types of orbits remain practically unperturbed by the shifting on the value of the orbital energy. For larger values of the energy $(\widehat{C} > 0.01)$ however, the percentages of escaping orbits exhibit a small divergence. Specifically, the rate of orbits escaping through $L_1$ slightly decreases reaching about 41\% at $\widehat{C} = 0.1$, while that of orbits escaping through $L_2$ continues the monotone behavior. It should be emphasized that the divergence of the percentages of escaping orbits observed in the phase space is much smaller than that found to exist in the configuration space. Once more the rate of trapped chaotic orbits remains always less than 0.5\%. Therefore, one may reasonably deduce that in the phase space the increase in the value of the energy does not influence significantly the orbital content of the dynamical system, thus leaving the percentages of the orbits almost the same throughout. In addition, since that the rates of escaping orbits remain unperturbed within the energy range, we may say that the two exit channels can be considered almost equiprobable in the phase space.

\subsection{An overview analysis}
\label{over}

\begin{figure*}[!tH]
\centering
\resizebox{0.9\hsize}{!}{\includegraphics{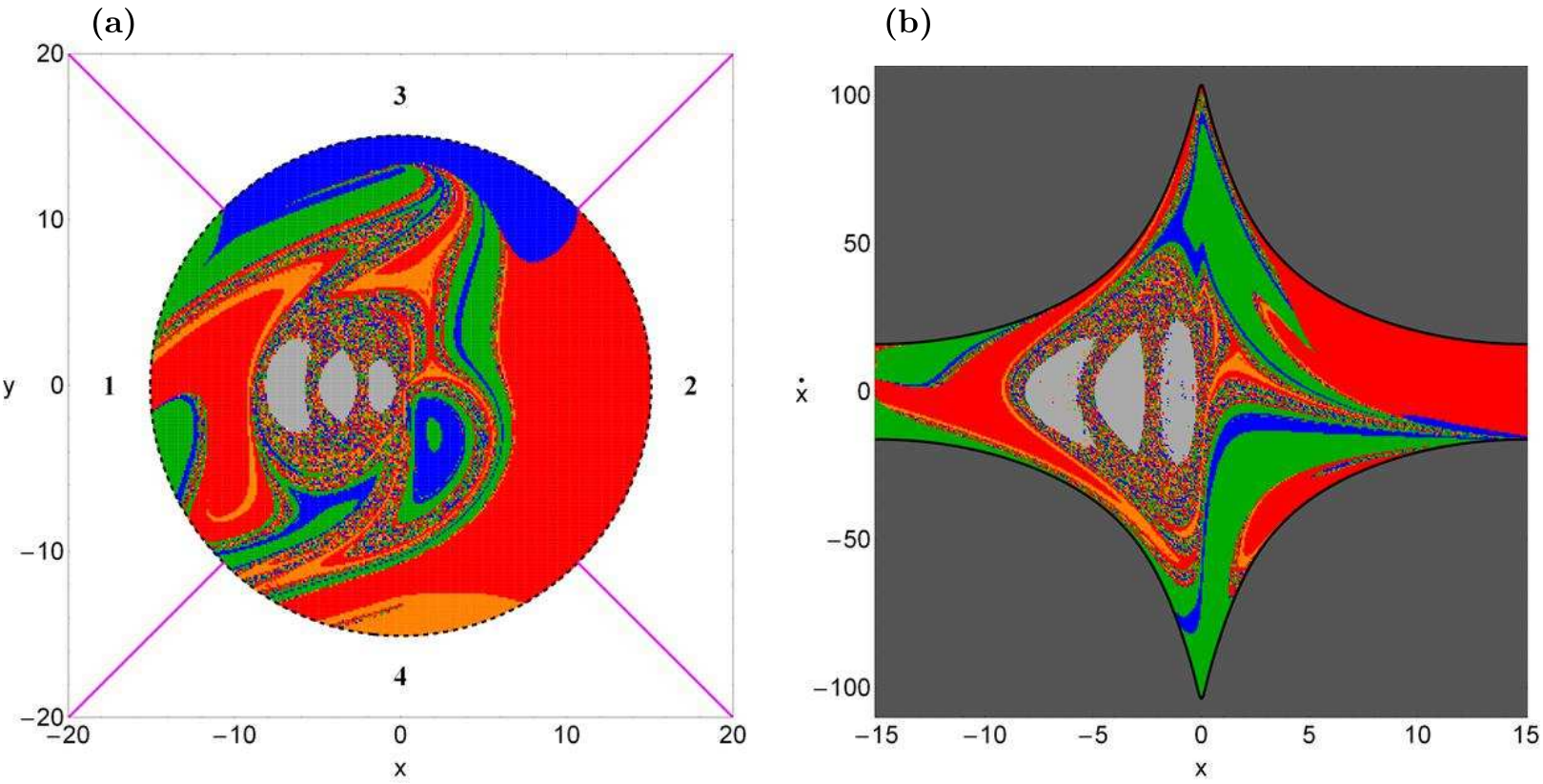}}
\caption{Orbital structure of the (a-left): configuration $(x,y)$ space and (b-right): phase $(x,\dot{x})$ space for $E = E(L_4) = -405.5861728613682$. The basins of the four sectors are: sector 1 (green), sector 2 (red), sector 3 (blue), sector 4 (orange). Light gray areas represent stability islands of regular non-escaping orbits, while initial conditions of trapped chaotic orbits are marked in white. The magenta straight lines delimit the boundaries of the four sectors.}
\label{ll}
\end{figure*}

\begin{figure*}[!tH]
\centering
\resizebox{0.9\hsize}{!}{\includegraphics{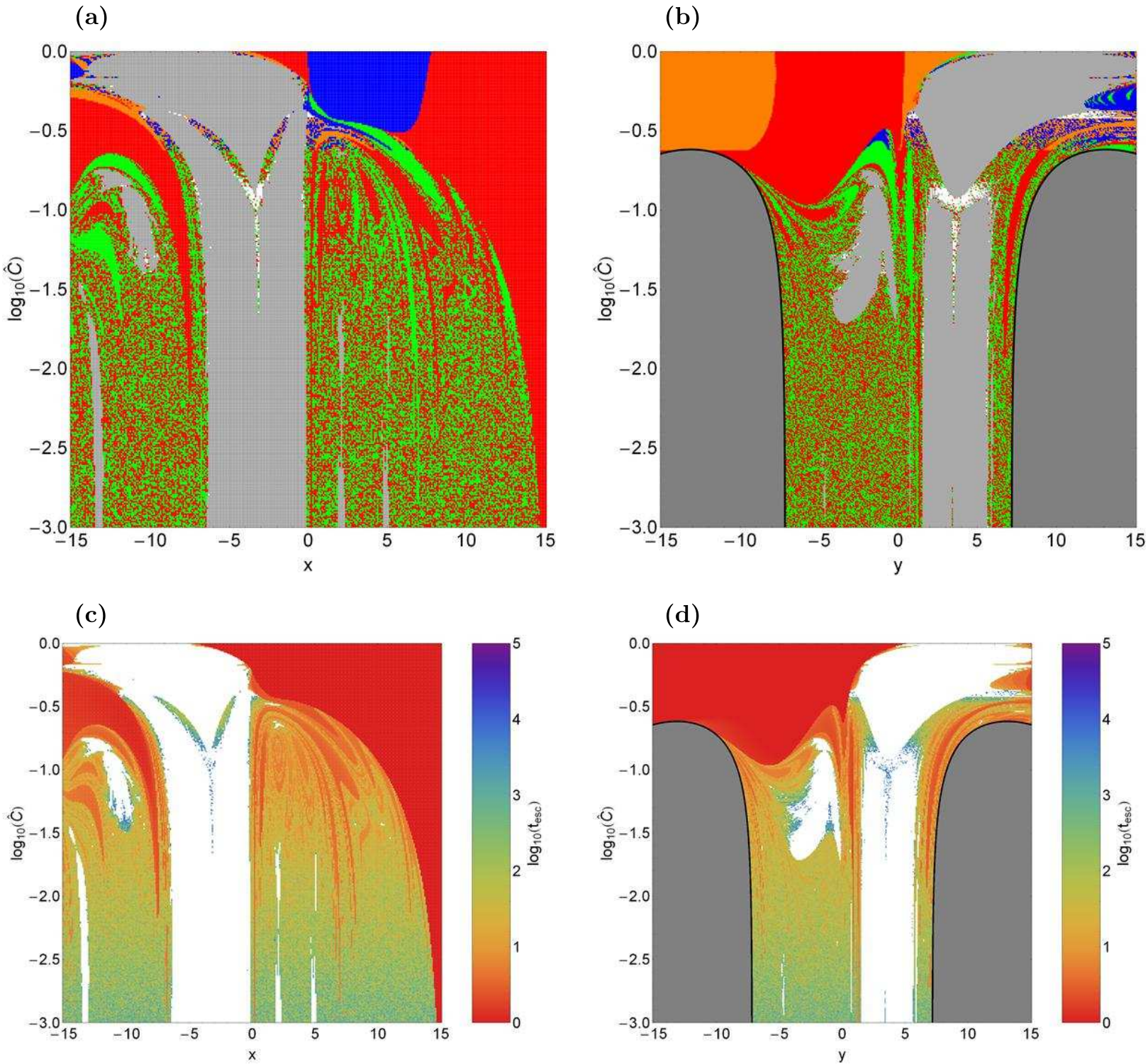}}
\caption{Orbital structure of the (a-upper left): $(x,\widehat{C})$-plane; and (b-upper right): $(y,\widehat{C})$-plane; (c-lower left and d-lower right): the distribution of the corresponding escape times of the orbits. The color code is exactly the same as in Fig. \ref{ll}.}
\label{xyc}
\end{figure*}

\begin{figure*}[!tH]
\centering
\resizebox{0.9\hsize}{!}{\includegraphics{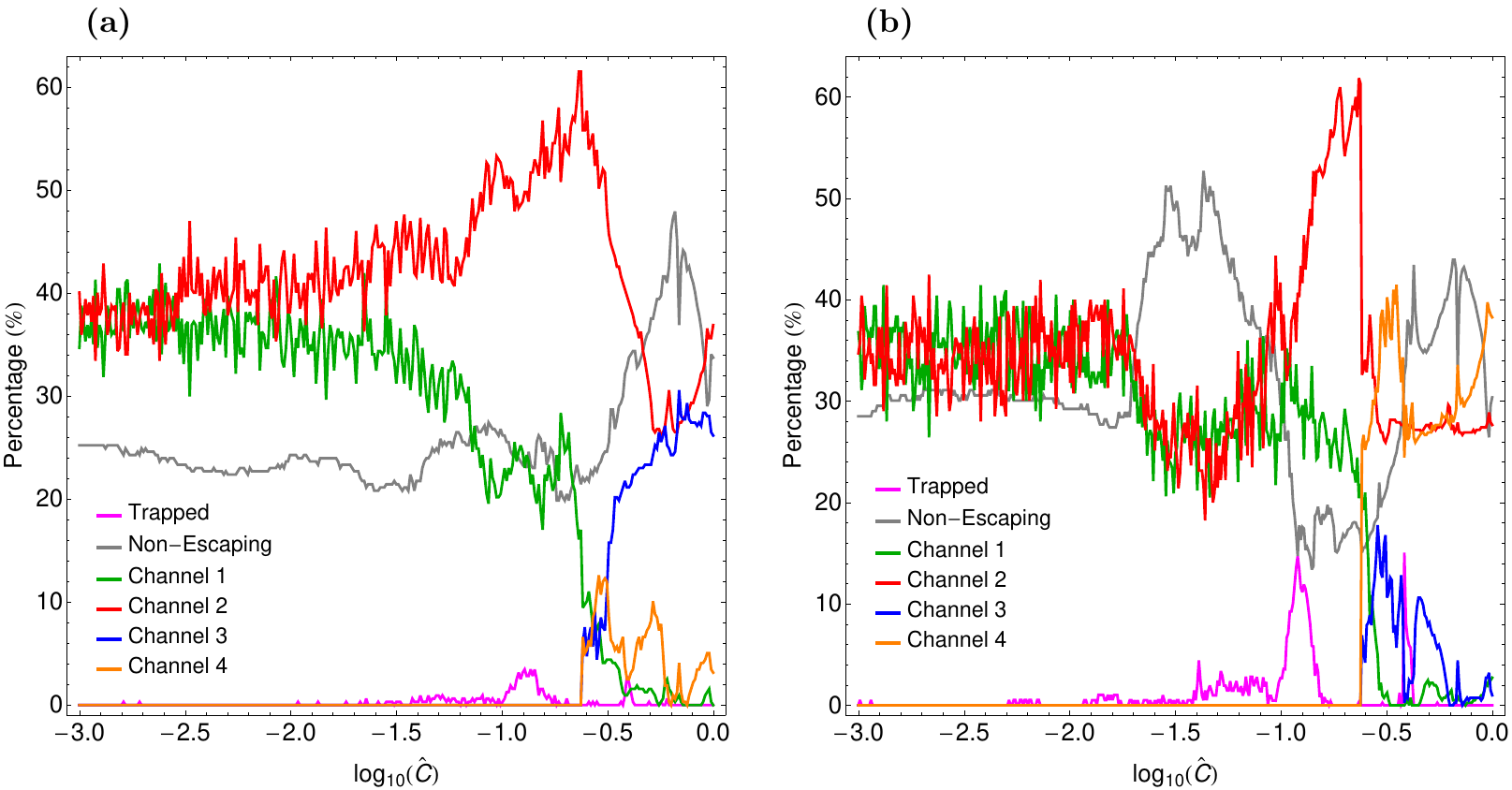}}
\caption{Evolution of the percentages of escaping, non-escaping regular and trapped chaotic orbits on the (a-left): $(x,\widehat{C})$-plane and (b-right): $(y,\widehat{C})$-plane as a function of the dimensionless energy parameter $\widehat{C}$.}
\label{percs3}
\end{figure*}

\begin{figure*}[!tH]
\centering
\resizebox{0.7\hsize}{!}{\includegraphics{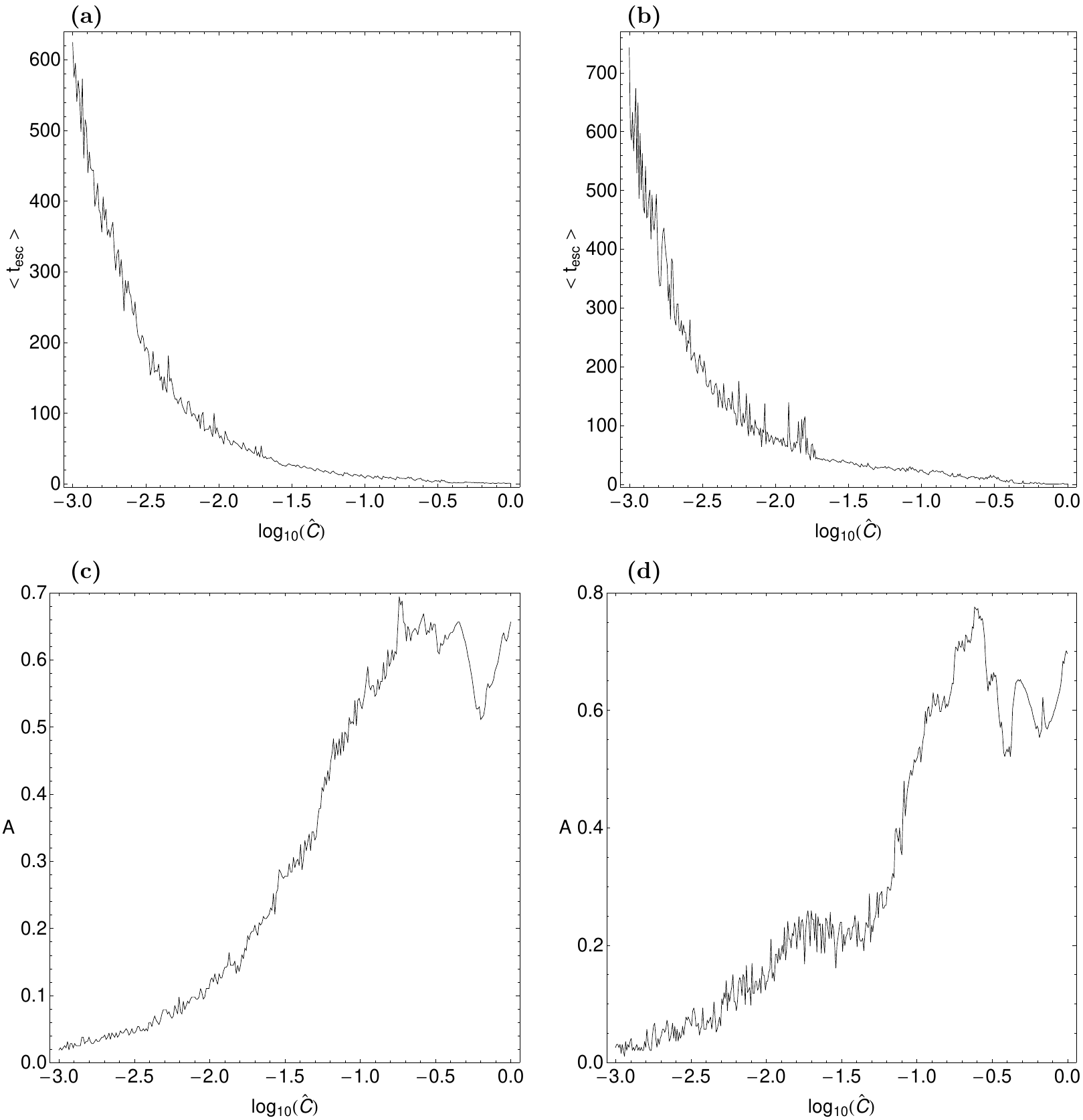}}
\caption{(a-b): The average escape time of orbits $< t_{\rm esc} >$ and (c-d): the percentage of the total area $A$ of the planes covered by the escape basins as a function of the dimensionless energy parameter $\widehat{C}$.}
\label{stats}
\end{figure*}

The color-coded grids in the configuration $(x,y)$ as well as in the phase $(x,\dot{x})$ space provide sufficient information on the phase space mixing however for only a fixed value of the Jacobi constant. H\'{e}non back in the late 60s [\citealp{H69}], introduced a new type of plane which can provide information not only about stability and chaotic regions but also about areas of trapped and escaping orbits using the section $y = \dot{x} = 0$, $\dot{y} > 0$ (see also [\citealp{BBS08}]). In other words, all the orbits of the stars of the quasar galaxy system are launched from the $x$-axis with $x = x_0$, parallel to the $y$-axis $(y = 0)$. Consequently, in contrast to the previously discussed types of planes, only orbits with pericenters on the $x$-axis are included and therefore, the value of the dimensionless energy parameter $\widehat{C}$ can be used as an ordinate. In this way, we can monitor how the energy influences the overall orbital structure of our dynamical system using a continuous spectrum of energy values rather than few discrete energy levels. We decided to explore the energy range when $\widehat{C} \in [0.001,1]$.

Here it should be pointed out that for $E > E(L_4) = -405.5861728613682$ the forbidden regions disappear and all the configuration $(x,y)$ space is available for motion. Therefore for $E > E(L_4)$ we need to introduce new escape criteria. In particular, we divide the $(x,y)$ plane into four sectors according to a polar angle $\theta$ which starts counting from the positive part of the $x$-axis $(x > 0, y = 0)$ using the approach followed in [\citealp{dAT14}]. So, considering that $\theta \in [0^{\circ},360^{\circ}]$, we have the first sector for $135^{\circ} \leq \theta < 225^{\circ}$, the second sector for $\theta \geq 315^{\circ}$ or $\theta < 45^{\circ}$, the third sector for $45^{\circ} \leq \theta < 135^{\circ}$ and the fourth sector for $255^{\circ} \leq \theta < 315^{\circ}$, respectively. We define the sectors in such a way so that sectors 1 and 2 to correspond to the previous escape channels 1 and 2, respectively. In Fig. \ref{ll}a we present the structure of the configuration $(x,y)$ plane for $E = E(L_4)$ where the initial conditions of the escaping orbits are colored according to the four sectors (the magenta lines denote the limits of each sector). It is seen that most of the orbits escape through sectors 1 and 2, while the basins corresponding to sectors 3 and 4 are smaller. So, despite the fact that the ZVC is not so restrictive when $E > E(L_4)$, escape occurs mostly to the right and to the left of the scattering region. In particular, for $E = E(L_4)$, 46\% of the total integrated initial conditions belongs to the basin of the second sector, 21.77\% belongs to the basin of the first sector, 17.68\% belongs to the basin of the third sector, and only 11.25\% belongs to the basin of the fourth sector. The remaining 3.3\% of the solutions belong to the bounded basins. For $E = E(L_4)$ we observe the presence of three distinct stability islands. It is also seen that in the phase space shown in Fig. \ref{ll}b there are still forbidden regions even for $E > E(L_4)$.

In Fig. \ref{xyc}a we present the orbital structure of the $(x,\widehat{C})$-plane when $\widehat{C} \in [0.001,1]$, while in Fig. \ref{xyc}c the distribution of the corresponding escape times of orbits is depicted. We observe that for relatively low energy values $(0.001 < \widehat{C} < 0.01)$ the interior region is highly fractal, while basins of escape are located only at the outer parts of the $(x,\widehat{C})$-plane, that is outside $L_1$ and $L_2$ (or in other words in the exterior region). Furthermore, we can identify the presence of one main stability island of retrograde $(x_0 < 0)$ non-escaping regular orbits and two stability islands of prograde $(x_0 > 0)$ regular motion. For larger values of energy $\widehat{C} > 0.01$ however, the structure of the $(x,\widehat{C})$-plane changes drastically and the most important differences are the following: (i) several basins of escape are formed inside the fractal escape region, (ii) one more stability island of retrograde orbits emerges. It should be pointed out that in the blow-ups of the diagram several additional extremely tiny islands of stability have been identified\footnote{An infinite number of regions of (stable) quasi-periodic (or small scale chaotic) motion is expected from classical chaos theory.}, (iii) at high enough energy levels $(\widehat{C} > 0.1)$ the basin of escape corresponding to exit 2 or sector 2 cover the vast majority of the grid, (iv) the extent of the main island of regular motion grows rapidly and for $\widehat{C} > 0.1$ the position of initial conditions of non-escaping regular orbits exceed the interior region $(x_0 < - r_L)$.

In order to obtain a more complete view on how the total orbital energy influences the nature of orbits in our quasar galaxy model, we follow a similar numerical approach to that explained before but now all orbits of stars are initiated from the vertical $y$-axis with $y = y_0$. In particular, this time we use the section $x = \dot{y} = 0$, $\dot{x} > 0$, launching orbits parallel to the $x$-axis. This allow us to construct again a 2D plane in which the $y$ coordinate of orbits is the abscissa, while the logarithmic value of the energy $\log_{10}(\widehat{C})$ is the ordinate. The orbital structure of the $(y,\widehat{C})$-plane when $\widehat{C} \in [0.001,1]$ is shown in Fig. \ref{xyc}b. The black solid line is the limiting curve which distinguishes between regions of allowed and forbidden motion and is defined as
\begin{equation}
f_L(y,\widehat{C}) = \Phi_{\rm eff}(0,y) = E.
\label{zvc2}
\end{equation}
A very complicated orbital structure is reveled in the $(y,\widehat{C})$-plane which however, in general terms, is very similar with that of the $(x,\widehat{C})$-plane. One may observe that for $E > E(L_4)$ the forbidden regions of motion around $L_4$ and $L_5$ completely disappear. In fact the value $E(L_4) = E(L_5)$ is another critical energy level. It is seen that above that critical value the left part $(y_0 < 0)$ of the plane is occupied almost completely with initial conditions of orbits that escape to sectors 2 or 4, while the right part of the same plane on the other hand, contains a mixture of initial conditions of escaping and non-escaping orbits.

It is evident from the results presented in Figs. \ref{xyc}(c-d) that the escape times of the orbits are strongly correlated with the escape basins. In addition, one may conclude that the smallest escape periods correspond to orbits with initial conditions inside the escape basins, while orbits initiated in the fractal regions of the planes or near the boundaries of stability islands have the highest escape rates. In both types of planes the escape times of orbits are significantly reduced with increasing energy. Combining all the numerical outcomes presented in Figs. \ref{xyc}(a-d) we may say that the key factor that determines and controls the escape times of the orbits is the value of the orbital energy (the higher the energy level the shorter the escape rates), while the fractality of the basin boundaries varies strongly both as a function of the energy and of the spatial variable.

The following Fig. \ref{percs3}(a-b) shows the evolution of the percentages of the different types of orbits on the $(x,\widehat{C})$ and $(y,\widehat{C})$ planes as a function of the dimensionless energy parameter $\widehat{C}$. We see in Fig. \ref{percs3}a that for $\widehat{C} < 0.1$ the percentages of escaping orbits through channels 1 and 2 exhibit similar fluctuations, while for larger energy levels they start to diverge. In particular, the rate of escapers through exit 2 increases, while on the other hand the rate of escapers through exit 1 decreases. The portion of escaping orbits through sectors 3 and 4 is smaller, while in general terms the percentage of non-escaping regular orbits drops, although for some energy intervals the amount of regular orbits presents some peaks. In the same vein one may observe in Fig. \ref{percs3}b that the percentages of escaping orbits through exits 1 and 2 identically evolve for $\widehat{C} < 0.1$. For higher values of the energy the rates diverge following completely different paths. We also see that each sector dominates for different range of the energy. It should be pointed out that for $\widehat{C} > 0.01$ the percentage of non-escaping ordered orbits exhibits several sudden peaks.

The evolution of the average value of the escape time $< t_{\rm esc} >$ of orbits as a function of the dimensionless energy parameter is given in Fig. \ref{stats}(a-b) for the $(x,\widehat{C})$ and $(y,\widehat{C})$ planes, respectively. It is seen, that for low values of energy, just above the escape value, the average escape time of orbits is about 600 time units, however it reduces rapidly tending asymptotically to zero which refers to orbits that escape almost immediately from the system. We feel it is important to justify this behaviour of the escape time. As the value of the Jacobi constant increases the escape channels (which are of course symmetrical) become more and more wide and therefore, orbits need less and less time in order to find one of the openings in the open ZVC and eventually escape from the system. This geometrical feature explains why for low values of energy orbits consume large time periods wandering inside the open ZVC until they eventually locate one of the two exits and escape to infinity. Finally Fig. \ref{stats}(c-d) shows the evolution of the percentage of the total area $(A)$ on the $(x,\widehat{C})$ and $(y,\widehat{C})$ planes corresponding to basins of escape, as a function of the dimensionless energy parameter. It is seen that for low values of the energy both types of planes are highly fractal. However, as we proceed to higher energy levels the degree of fractalization reduces and the area corresponding to basins of escape start to grow rapidly. Eventually, at very high energy levels $(\widehat{C} = 1)$ the fractal domains are confined and therefore the well formed basins of escape occupy more than 70\% of the entire planes.

\section{Discussion and conclusions}
\label{disc}

The aim of this work was to shed some light to the trapped or escaping character of orbits in a dynamical gravitational model describing a quasar host galaxy with a disk and a central, dense and spherically symmetric nucleus. The particular dynamical model we used for describing the host galaxy was proposed in Paper I. As far as we know, this is the first time that the escape process of stars in a quasar active galaxy is systematically investigated in such detail.

The Jacobi integral of motion has a critical threshold value (or escape energy) which determines if the escape process is possible or not. In particular, for energies smaller than the threshold value, the ZVC is closed and it is certainly true that escape is impossible. For energy levels larger than the escape energy however, the ZVC opens and two exit channels appear through which the stars can escape to infinity. Here it should be emphasized, that if a star does have energy larger than the escape value, there is no guarantee that the star will certainly escape from the system and even if escape does occur, the time required for an orbit to transit through an exit and hence escape to infinity may be very long compared with the natural escaping time. We managed to distinguish between ordered/chaotic and trapped/escaping orbits and we also located the basins of escape leading to different exit channels, finding correlations with the corresponding escape times of the orbits. Our extensive and thorough numerical investigation strongly suggests, that the overall escape mechanism is a very complicated procedure and very dependent on the value of the Jacobi integral.

We defined for several values of the Jacobi integral, dense uniform grids of $1024 \times 1024$ initial conditions regularly distributed in the area allowed by the energy level on the configuration and phase space, respectively and then we identified regions of order/chaos and bound/escape. For the numerical integration of the orbits in each type of grid, we needed about between 0.5 hour and 12 days of CPU time on a Pentium Dual-Core 2.2 GHz PC, depending on the escape rates of orbits in each case. For each initial condition, the maximum time of the numerical integration was set to be equal to $10^4$ time units. However, when a star escaped the numerical integration was effectively ended and proceeded to the next available initial condition.

The present article provides quantitative information regarding the escape dynamics in a quasar galaxy system. The main numerical results of our research can be summarized as follows:
\begin{enumerate}
 \item A strong correlation between the value of the energy (Jacobi integral) and the extent of the basins of escape was found to exist. Indeed, for very low energy levels the structure of both the configuration and the phase space exhibits a large degree of fractalization and therefore the majority of orbits escape choosing randomly exit channels. As the value of the energy increases however, the structure becomes less and less fractal and several basins of escape emerge. The extent of these basins of escape is more prominent at relatively high energy levels.
 \item In all examined cases, areas of bounded motion (ordered and chaotic) and regions of initial conditions leading to escape in a given direction (basins of escape), were found to exist in both the configuration and the phase space. The several escape basins are very intricately interwoven and they appear either as well-defined broad regions or thin elongated bands. Regions of bounded orbits first and foremost correspond to stability islands of regular orbits where a third integral of motion is present.
 \item We observed that that the escape times of orbits are directly linked to the basins of escape. In particular, inside the basins of escape as well as relatively away from the fractal domains, the shortest escape rates of the orbits had been measured. On the other hand, the longest escape periods correspond to initial conditions of orbits either near the boundaries between the escape basins or in the vicinity of the stability islands.
 \item It was detected that in many cases the escape process is highly sensitive dependent on the initial conditions, which means that a minor change in the initial conditions of an orbit lead the star to escape through the opposite exit channel. These regions are completely intertwined with respect to each other (fractal structure) and are mainly located in the vicinity of stability islands. This sensitivity towards slight changes in the initial conditions in the fractal regions implies, that it is impossible to predict through which exit the particle will escape.
 \item Our numerical computations revealed that for energy levels slightly above the escape energy the majority of the escaping orbits have considerable long escape rates (or escape periods), while as we proceed to higher energies the proportion of fast escaping orbits increases significantly. This behavior was justified, by taken into account a geometrical feature of the ZVC. Specifically, for low values of energy the width of the exit channels is too small therefore, stellar orbits consume large time periods wandering inside the open ZVC until they eventually locate one of the two exits and escape to infinity.
\end{enumerate}

Judging by the detailed and novel outcomes we may say that our task has been successfully completed. We hope that the present numerical analysis and the corresponding results to be useful in the field of escape dynamics in active galaxy systems. The outcomes as well as the conclusions of the present research are considered, as an initial effort and also as a promising step in the task of understanding the escape mechanism of orbits in active host galaxies. Taking into account that our results are encouraging, it is in our future plans to modify properly our dynamical model in order to expand our investigation into three dimensions and explore the entire six-dimensional phase space.

\section*{Acknowledgment}

The author would like to thank the two anonymous referees for the careful reading of the manuscript and for all the apt suggestions and comments which allowed us to improve both the quality and the clarity of the paper.

\section*{Compliance with Ethical Standards}

\begin{itemize}
  \item Funding: The author states that he has not received any research grants.
  \item Conflict of interest: The author declares that he has no conflict of interest.
\end{itemize}

\end{document}